\documentclass[12pt,preprint]{aastex}
\shorttitle{Radio emission from KISSR\,1494}
\shortauthors{Kharb et al.}
\begin{document}
\title{VLBI Imaging of the Double Peaked Emission Line Seyfert KISSR\,1494}
\author{P. Kharb}
\affil{Indian Institute of Astrophysics, II Block, Koramangala, Bangalore 560034, India}
\email{kharb@iiap.res.in}
\author{M. Das}
\affil{Indian Institute of Astrophysics, II Block, Koramangala, Bangalore 560034, India}
\author{Z. Paragi}
\affil{Joint Institute for VLBI in Europe, Postbus 2, 7990 AA Dwingeloo, the Netherlands}
\author{S. Subramanian}
\affil{Indian Institute of Astrophysics, II Block, Koramangala, Bangalore 560034, India}
\author{L. P. Chitta}
\affil{Indian Institute of Astrophysics, II Block, Koramangala, Bangalore 560034, India}
\begin{abstract}
We present here the results from dual-frequency phase-referenced VLBI observations of the Seyfert galaxy KISSR\,1494, which exhibits double peaked emission lines in its SDSS spectrum. We detect a single radio component at 1.6 GHz, but not at 5~GHz implying a spectral index steeper than $-1.5\pm0.5$ ($S_\nu\propto\nu^\alpha$). The high brightness temperature of the radio component ($\sim1.4\times10^7$~K) and the steep radio spectrum support a non-thermal synchrotron origin. A crude estimate of the black hole mass derived from the $M_{BH}-\sigma_{\star}$ relation is $\sim1.4\pm1.0\times10^8~M_{\sun}$; it is accreting at an Eddington rate of $\sim0.02$. The radio data are consistent with either the radio emission coming from the parsec-scale base of a synchrotron wind originating in the magnetised corona above the accretion disk, or from the inner ionised edge of the accretion disk or torus. In the former case, the narrow line region (NLR) clouds may form a part of the broad outflow, while in the latter, the NLR clouds may form a part of an extended disk beyond the torus. The radio and NLR emission may also be decoupled so that the radio emission originates in an outflow while the NLR is in a disk, and {\it vice versa}. While with the present data, it is not possible to clearly distinguish between these  scenarios, there appears to be greater circumstantial evidence supporting the coronal wind picture in KISSR\,1494. From the kiloparsec-scale radio emission, the time-averaged kinetic power of this outflow is estimated to be $Q\approx1.5\times10^{42}$~erg~s$^{-1}$, which is typical of radio outflows in low-luminosity AGN. This supports the idea that radio ``jets'' and outflowing coronal winds are indistinguishable in Seyfert galaxies. 
\end{abstract}
\keywords{Seyfert galaxies: general --- Seyfert galaxies: individual (KISSR\,1494)}

\section{Introduction}
\label{secintro}
{It is now widely believed that all luminous galaxies host supermassive black holes (SMBH, $M_{BH}\sim10^6-10^9~M_\sun$) in their centres \citep[e.g.,][]{Kormendy13}. Since mergers are an essential part of the evolutionary process of a galaxy \citep[e.g.,][]{Barnes92}, the presence of multiple SMBH in galactic nuclei are expected \citep[e.g.,][]{Volonteri03}. However, so far, there are only 18 confirmed dual/binary accretion-powered active galactic nuclei (AGN) known in the literature \citep{Deane14}. Only three of these systems have black hole (BH) separations $<1$~kiloparsec. These are 0402+379 \citep[BH separation $\sim$7 parsec;][]{Rodriguez06}, NGC\,6240 \citep[BH separation $\sim$70 parsec;][]{Komossa03} and NGC\,3393 \citep[BH separation $\sim$150 parsec;][]{Fabbiano11}. The {dual} AGN candidate (with a BH separation of $\sim$77 parsec) identified by \citet{Gitti13} in the galaxy cluster RBS\,797 needs to be confirmed by multi-frequency radio imaging which should ideally detect two flat spectrum radio cores \citep[e.g., see][]{BurkeSpolaor11}. However, as pointed out by \citet{Wrobel14}, single young AGN may show double radio structures with relatively compact flat spectrum ``hot spots" that may be mistaken for dual AGN cores in sensitivity-limited observations.

Almost all of the dual/binary AGN candidates \citep[see][]{Deane14} reside in merger remnants or elliptical galaxies, with the rare exception of NGC\,3393 which resides in a spiral galaxy \citep{Fabbiano11}. The merger process, which creates large elliptical galaxies \citep[e.g.,][]{Steinmetz02}, results in gas infall, massive star formation and the formation of supermassive BH binaries. Spiral galaxies, on the other hand, undergo minor mergers (e.g., satellite accretion) which result in larger bulges and intact disks \citep{Aguerri01}: binary SMBH must therefore be rarer in spiral galaxies. 

There are attempts in the literature to identify a greater number of dual/binary AGN. One of the adopted techniques is via the identification of AGN with double-peaked emission lines (DPAGN), with the assumption that the split in the emission lines is a consequence of two BHs carrying along with them their respective narrow/broad emission line regions (NLR/BLR) having different characteristic velocities. While some DPAGN have indeed turned out to be dual AGN \citep[e.g.,][]{Fu11}, it is becoming apparent that this technique is not necessarily more efficient in detecting binary BHs than other techniques like identifying sources with S- or X-shaped jets. For instance, \citet{Fu12} discovered only two binary AGN in a sample of 42 DPAGN. As is becoming clear in more and more sources, double-peaked emission lines could result from radio outflows and jet-medium interaction \citep[e.g.,][]{Rosario10,Fischer11,Gabanyi14}, or rotating gaseous disks \citep[e.g.,][]{Shen11,Smith12}. Nevertheless, the search for binary black holes in DPAGN goes on.

Both the Chandra X-ray Observatory and the Hubble Space Telescope (HST) have been instrumental in the identification and confirmation of dual AGN with kiloparsec-scale black hole separations \citep[e.g.,][]{Komossa03,Fabbiano11}. However, \citet{Deane14} have noted that $>60\%$ of the sub-10\,kpc dual/triple AGN candidates have associated radio emission: BH activity and radio jet formation are enhanced in systems with two or more BHs, which are a consequence of mergers. The technique of Very Long Baseline Interferometry (VLBI) is essential to identify and confirm radio-emitting binary AGN with parsec-scale BH separations: multiple frequencies must be used to check if the core spectra are flat/inverted. 

Seyfert galaxies have traditionally been identified as ``radio-quiet'' AGN \citep[however see][]{HoPeng01,Kharb14a}. Therefore, the detection of binary radio-emitting AGN in a Seyfert nucleus is doubly challenging. In the current paper, we present results from VLBI observations of a Seyfert galaxy with double peaked emission lines in its SDSS spectrum. However, before discussing this source, it is worth discussing the complex nature of radio emission in Seyfert galaxies.

Unlike the outflows in powerful radio galaxies, Seyfert outflows are a poorly understood phenomenon. Only sensitive radio observations at low radio frequencies reveal the presence of radio structures with typical extents $\lesssim$10 kiloparsec in Seyfert galaxies. \citet{Gallimore06} and \citet{Singh14} have found that $\gtrsim40\%$ of Seyferts belonging to large (complete or eclectic) samples exhibit kiloparsec-scale radio structures when observed with sensitive radio arrays (e.g., VLA in the D-array configuration). It is unclear {\it how} or {\it where} these radio outflows are generated. The contribution of the AGN versus star-formation to the radio emission is also debatable \citep[e.g.,][]{Baum93,Hota06}. \citet{Malzac01} and \citet{Markoff05} have argued the case for Seyfert outflows being outflowing accretion-disk coronae. High resolution VLBI observations on the other hand, have detected the presence of parsec-scale jets in a majority of radio-bright Seyferts \citep[e.g.,][]{Nagar05,Kharb10a,Kharb14a,Mezcua14}. The VLBI cores however do not always exhibit flat/inverted {spectra} and could {have} steep {spectra} instead \citep[e.g.,][]{Roy00,Orienti10,Kharb10a,Bontempi12,Panessa13}, unlike the unresolved bases of relativistic jets in powerful radio galaxies. 

We present here the results from multi-frequency VLBA observations of the Seyfert 2 galaxy KISSR\,1494 belonging to the KPNO Internal Spectroscopic Survey (KISS) of spiral/disk emission line galaxies \citep{Wegner03}.
}
\begin{figure}[t]
\centerline{
\includegraphics[width=10cm,trim=1.8cm 0cm 0cm 0cm]{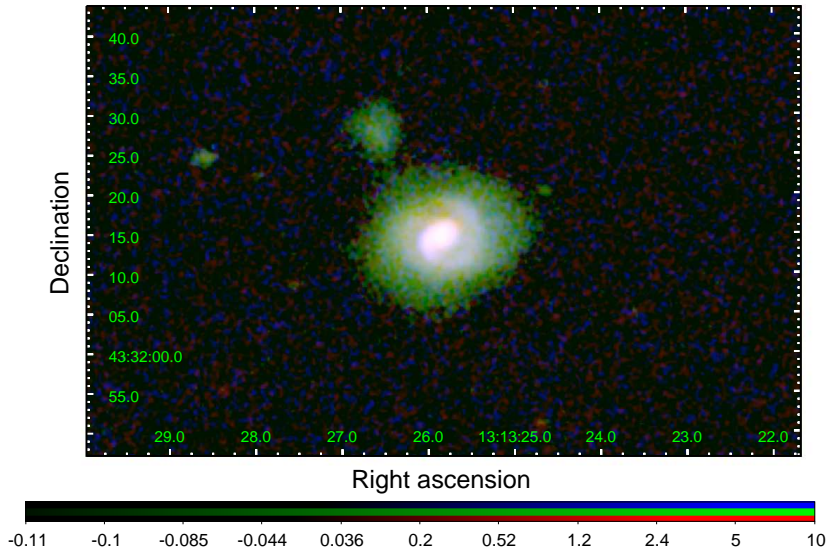}}
\caption{RGB image of KISSR\,1494 made using $z$ (central $\lambda = $9134\AA), $r$ (= 6231\AA) and $u$ (=3543\AA) filters of the SDSS.}
\label{figrgb}
\end{figure}
KISS is a nearly complete sample of emission-line galaxies over a well-defined volume in space. The survey was carried out using the 2.4m MDM telescope on Kitt Peak and contains 351 galaxies. We searched through the KISS sample for signatures of double peaked emission line AGN. We examined the SDSS\footnote{Sloan Digital Sky Survey \url https://www.sdss3.org/dr10} spectra of those galaxies which had been classified as either Seyfert 1, Seyfert 2, or LINER\footnote{Low-Ionization Nuclear Emission-line Region galaxies} (a total of 72 galaxies), and identified six ({\it i.e.}, 8\%) as having double peaks in their emission line spectra. Only three of these, viz., KISSR\,434, KISSR\,1219 and KISSR\,1494, had been detected in the VLA FIRST\footnote{Faint Images of the Radio Sky at Twenty-Centimeters \url http://sundog.stsci.edu} and NVSS\footnote{The NRAO VLA Sky Survey \url http://www.cv.nrao.edu/nvss} surveys at 1.4~GHz (resolution $\sim5.4\arcsec$ and $\sim45\arcsec$, respectively). KISSR\,1494 was the brightest of these three with an integrated flux density of $\sim23$~mJy and $\sim$25~mJy in the FIRST and NVSS images, respectively. 
{The focus on KISSR\,1494 in this paper therefore merely stems from the fact that it is bright enough to be observed with VLBI.}

\begin{figure}[t]
\centerline{
\includegraphics[width=10cm]{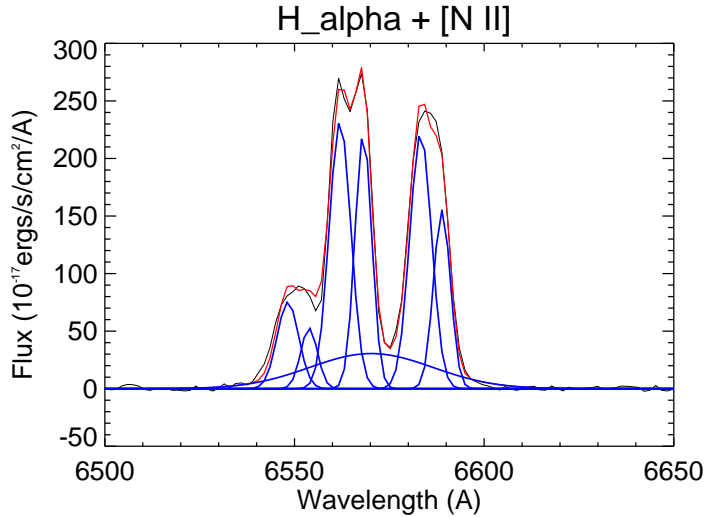}}
\caption{SDSS spectrum of KISSR\,1494 showing the dereddened double-peaked H$\alpha$+[N {\sc ii}] lines in black, Gaussian line fits in blue and the total spectrum in red.}
\label{figspectra1}
\end{figure}
\begin{figure}[t]
\centerline{
\includegraphics[width=16cm]{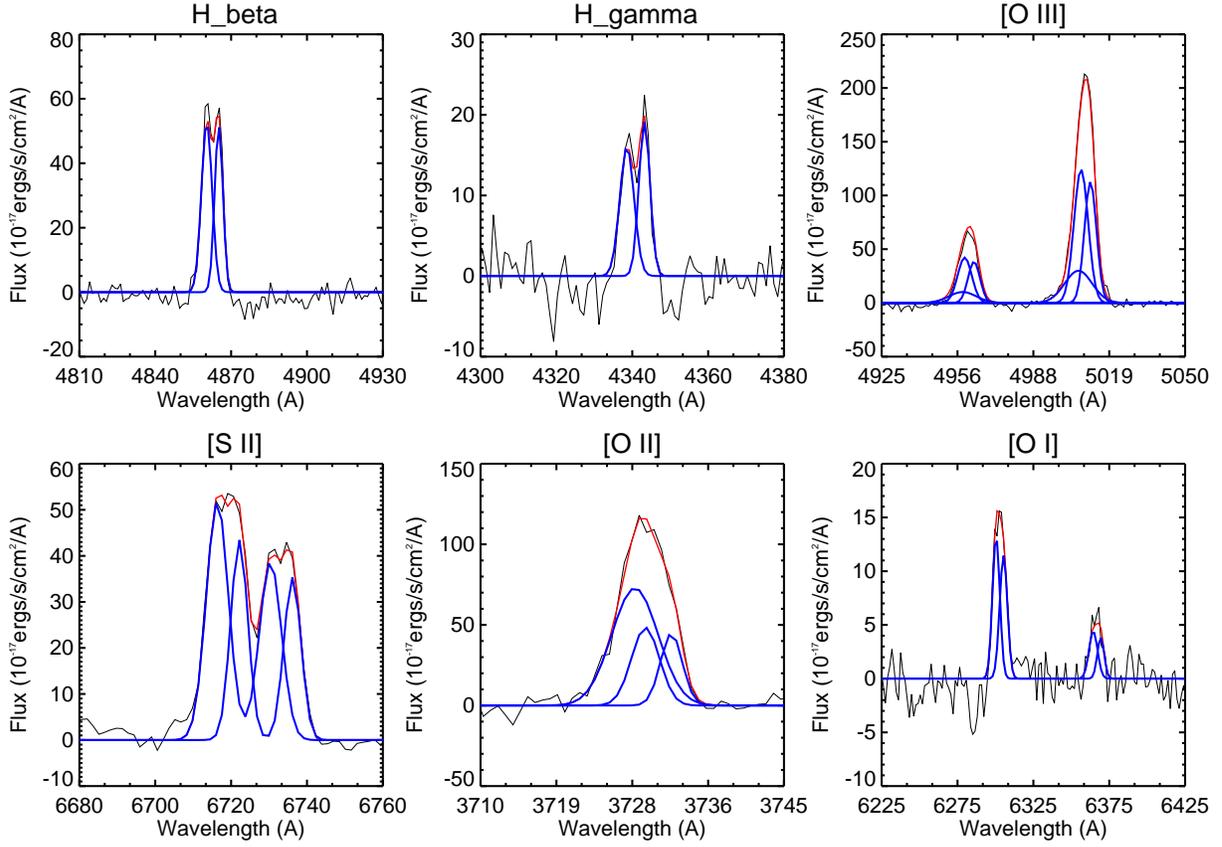}}
\caption{SDSS spectrum of KISSR\,1494 showing the dereddened double-peaked H$\beta$, H$\gamma$, [S {\sc ii}], [O {\sc i}] lines and the asymmetric [O {\sc ii}], [O {\sc iii}] lines in black. The Gaussian line fits are shown in blue and the total spectrum in red.}
\label{figspectra2}
\end{figure}

KISSR\,1494 has a high {\it IRAS} 100$\mu$m flux density ($\sim4.8$~Jy) and soft X-ray (0.1$-$2.4 keV) luminosity in the {\it ROSAT} sky survey \citep[$\mathrm{L_X}=1.05\times10^{42}$ erg~s$^{-1}$;][]{Stevenson02}. The X-ray luminosity was derived for an absorbed power-law model with $\Gamma=2.5$ and $\mathrm{N_H}$\,=\,Galactic\,+\,(2.0$\times10^{20}$)\,cm$^{-2}$. The SDSS image of KISSR\,1494 reveals a disky structure, a prominent bulge, and a one-armed spiral feature (Figure~\ref{figrgb}). A one-armed spiral indicates an $m=1$ instability which could arise due to counter-rotating gas/stars \citep[e.g.,][]{Zang78}. One explanation for such counter-rotating material could be the ingestion of small galaxy, small enough to not have destroyed the spiral structure of the larger galaxy. 

The SDSS spectrum of KISSR\,1494 shows several emission lines that are characteristic of AGN activity, the most prominent being H$\alpha$, H$\beta$ and [O {\sc iii}]\footnote{The SDSS spectra are acquired through a fiber of diameter $3\arcsec$, or 3.24~kiloparsec at the distance of KISSR\,1494}. {The H$\alpha$, H$\beta$ and H$\gamma$ lines} have distinct double peaks (Figures~\ref{figspectra1}, \ref{figspectra2}). As the two components of the narrow H$\alpha$, H$\beta$ lines have nearly equal strength, this source can be classified as an equal-peaked AGN \citep{Smith12}. Traces of double peaks are also observed in the [S {\sc ii}], [O {\sc i}] and [O {\sc ii}] lines. The [N {\sc ii}] and [O {\sc iii}] lines appear broad and asymmetric, possibly due to {blended} double peaks.

At the KISSR\,1494 redshift of $z=0.057446$ (luminosity distance, $D_L$ = 250 Megaparsec), 1~milliarcsec (mas) corresponds to a linear extent of 1.08 parsec for H$_0$ = 73~km~s$^{-1}$~Mpc$^{-1}$, $\Omega_{mat}$ = 0.27, $\Omega_{vac}$ =  0.73. The spectral index, $\alpha$, is defined throughout such that the flux density at frequency $\nu$ is $S_\nu\propto\nu^\alpha$.

\section{Observations and Data Reduction}
\subsection{Phase-referenced VLBI}
The observations were carried out with nine antennas of the Very Long Baseline Array (VLBA) at 1.55 and 4.98~GHz, on August 10, 2013 (Project ID: BK182); the Fort Davis antenna was not involved in the experiment. Fringes were not detected with the Pie Town antenna. Therefore, effectively only eight antennas of the VLBA were used for the observations. The data were acquired with an aggregate bit rate of 1024 Mbits~sec$^{-1}$ (2 polarization, 2 baseband channels, bandwidth 128~MHz, and a 2-bit sampling rate), in a phase-referencing mode. 1325+436 which is 2.53$\degr$ away from the source and has an x,~y positional uncertainty of 0.18,~0.28 mas, was used as the phase reference calibrator. 3C\,345 and 1315+415 were used as the fringe-finder and phase-check calibrator, respectively. A nodding cycle of 5 mins (2 mins on the phase calibrator and 3 mins on the target) was used for the observations, interspersed by two 5 min scans on 3C\,345 and two 3 mins scans on the phase-check calibrator (only a single scan of the latter at 5~GHz). The experiment lasted a total of three hours, with $\approx1.5$ hrs at each frequency. 

The data reduction was carried out using AIPS (Astronomical Image Processing System; version 31DEC12) following standard VLBA-specific tasks and procedures outlined in the AIPS cookbook\footnote{http://www.aips.nrao.edu/cook.html}. Los Alamos (LA), which was a stable antenna in the middle of the configuration, was used as the reference antenna for the calibration. The amplitude calibration was carried out using the procedure {\tt VLBACALA}, while the delay, rate, and phase calibration were carried out using the procedures {\tt VLBAPANG, VLBAMPCL} and {\tt VLBAFRGP}. The phase calibrator 1325+436 was iteratively imaged and self-calibrated on both phase and phase+amplitude using AIPS tasks {\tt IMAGR} and {\tt CALIB}. The images were then used as models to determine the amplitude and phase gains for the antennas. These gains were applied to the target and the final images were made using {\tt IMAGR}. A round of data-flagging was carried out using the task {\tt IBLED} on the source {\tt SPLIT} file, prior to making the images. The radio component detected at 1.6~GHz was offset from the centre of the image by 0.076$\arcsec$, 0.120$\arcsec$ in right ascension and declination. We ran the task {\tt UVFIX} on the source {\tt SPLIT} file to shift the source to the centre before producing the final image. We used the elliptical Gaussian-fitting task {\tt JMFIT} to obtain the component flux densities. 

\subsection{Emission-Line Fitting}
{To fit the emission lines in the SDSS spectrum of KISSR1494 and simultaneously obtain the stellar velocity dispersion from the absorption lines, we first separated the emission line spectrum from the underlying stellar continuum by the following steps. We corrected for reddening using SDSS tabulated E(B-V) values \citep{Schlegel98} and then used the {\tt pPXF} (Penalized Pixel-Fitting stellar kinematics extraction) code by \citet{Cappellari04} to obtain the best fit model for the underlying stellar population.
{We corrected the spectrum for the redshift of the galaxy as listed in SDSS dr10.}
The {\tt MILES} single stellar population models were used as templates of underlying stellar population as they cover a large range of metallicity (M/H $\sim-2.32$ to +0.22) and age (63 Myr to 17 Gyr). The stellar population contribution to the best fit template is $\sim$82\% and the remaining 18\% accounts for the contribution from other components like, the power law component of AGN and Fe II lines. The stellar velocity dispersion obtained for KISSR\,1494 from the absorption lines in the underlying stellar continuum is 206.6$\pm$8.2~km~s$^{-1}$.

The extracted pure emission line spectrum shows signatures of double peaks in prominent emission lines such as [S {\sc ii}], H$\alpha$, H$\beta$, H$\gamma$, [O {\sc i}], [O {\sc ii}], [O {\sc iii}]. 
{Our analysis used IDL programs that included the {\tt MPFIT} function for fitting Gaussians to emission lines. We first modelled each [S {\sc ii}] doublet line as two Gaussians having equal line widths in velocity space. The [S {\sc ii}] model was used as a template to fit the narrow [N {\sc ii}] doublet lines as well as H$\alpha$ and H$\beta$ narrow lines. This procedure has been described by \citet{Reines13}. The separation between the centroids of the [N {\sc ii}] narrow components was held fixed and the flux of [N {\sc ii}] $\lambda$6583 to [N {\sc ii}] $\lambda$6548 was fixed at the theoretical value of 2.96. With the inclusion of a broad H$\alpha$ component in addition to the two narrow components, the reduced $\chi^2$ improved from 1.49 to 1.09. For the [O {\sc iii}] doublet lines an additional third broad component  improved the fit by $\approx20\%$ ($\chi^2$ improved from 1.67 to 1.35). For the [O {\sc ii}] doublet lines however, the broad component could be a consequence of the two blended/unresolved components in one of the doublet lines. 
The central wavelength of the broad component of H$\alpha$ is redshifted by $7.23\pm0.65\AA$ while [O {\sc iii}] is blueshifted by $2.64\pm1.75\AA$. 
The velocity widths of the broad H$\alpha$ and [O {\sc iii}] components are 739.4$\pm$26.3~km~s$^{-1}$ and 338.6$\pm$53.0~km~s$^{-1}$, respectively. 
The larger width of the H$\alpha$ broad component compared to the [O {\sc iii}] broad component could signify a combination of Doppler broadening effects due to the black hole and an outflow, both in H$\alpha$ and the nearby [N {\sc ii}] doublet lines. The difference in the shifts of the central wavelengths of the broad components of [O {\sc iii}] and H$\alpha$ could also be attributable to the same effect. 
All the prominent lines along with their best fitting components} are shown in Figures~\ref{figspectra1}, \ref{figspectra2}; their details are noted in Table~\ref{tabprop}. 

\begin{figure}[t]
\centerline{
\includegraphics[width=10cm]{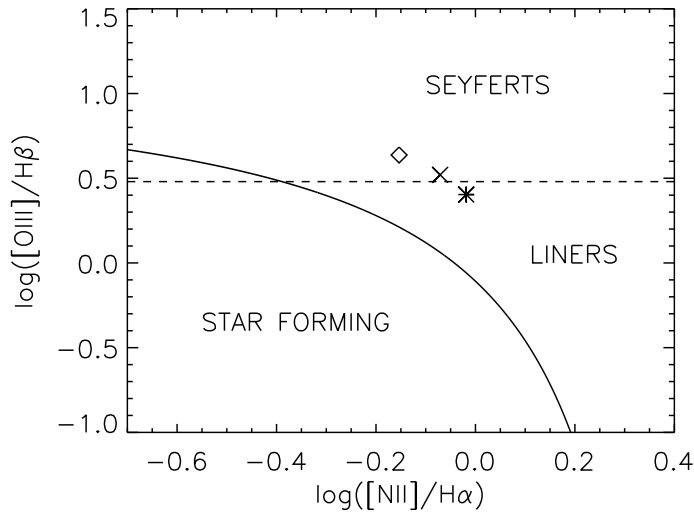}}
\caption{AGN diagnostic diagram for KISSR\,1494. The $\diamond$ and $\ast$ symbols represent the line ratios for the separate components of the double peaked emission lines, while the $\times$ symbol represents line ratios derived from the total fluxes. All three line ratios are consistent with a Seyfert nucleus given the $\lesssim$10\% errors in line-fitting. It is evident however that KISSR\,1494 lies close to the Seyfert-LINER boundary.}
\label{figagn}
\end{figure}

\section{Results}
The AGN diagnostic diagram for KISSR\,1494 confirms its Seyfert classification (Figure~\ref{figagn}). The solid line is the demarcation between star formation (HII regions) and active nuclei from \citet{Kewley01}. The dashed horizontal line is the separation between Seyfert and LINER nuclei \citep[e.g.,][]{Veilleux87}. We derived the line ratios log([O {\sc iii}], $\lambda$5007)/H$\beta$ and log([N {\sc ii}] $\lambda$6583)/H$\alpha$ in two ways: first, by taking the peaks separately and second, by combining all the emission as a mean value. Hence, Figure~\ref{figagn} has three points, with the mean line ratio being the central point in the plot (symbol $\times$). Keeping in mind that the errors in the line-fitting analysis can be $\lesssim$10\%, all three line ratios are consistent with a Seyfert nucleus. However, it is {also} evident from Figure~\ref{figagn} that KISSR\,1494 lies close to the Seyfert-LINER dividing line.

\subsection{Parsec-scale Radio Emission}
A single radio component was detected at 1.6~GHz (Figure~\ref{figcont1}), but not at 5~GHz. The radio component detected at 1.6~GHz has a peak intensity of $\sim360\,\mu$\,Jy~beam$^{-1}$, and an integrated flux density of $\sim$650\,$\mu$Jy in a purely naturally-weighted image (made with a {\tt ROBUST} parameter of +5 in AIPS task {\tt IMAGR}). The rms noise reached at 1.6~GHz was $\sim6.5\times10^{-5}$\,Jy~beam$^{-1}$, making this a 10$\sigma$ detection. At 5~GHz, where there was no source detection, a final rms noise of $\sim3.5\times10^{-5}$\,Jy~beam$^{-1}$ was reached. The deconvolved size of the radio component at 1.6~GHz was obtained from the purely naturally-weighted image using {the AIPS task} JMFIT. This size is $\sim7.5\times5$~mas, or $\sim8\times6$~parsec at the distance of the source. 

\begin{figure}[t]
\centerline{
\includegraphics[width=10cm,trim=1.8cm 0cm 0cm 0cm]{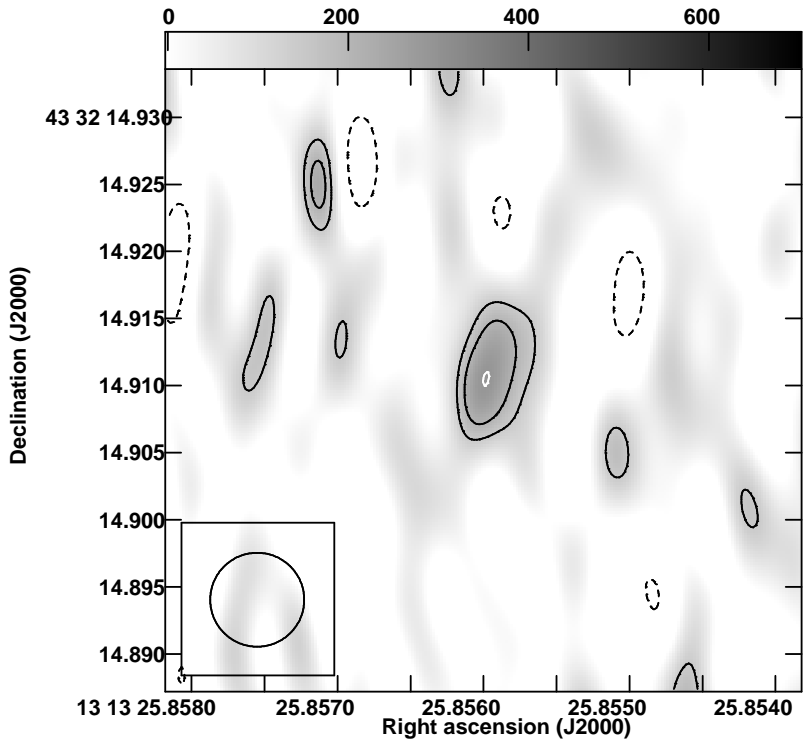}}
\caption{1.6~GHz contour image of KISSR\,1494. The contours are in percentage of the peak intensity and increase in steps of $\sqrt{2}$: the lowest contour and peak intensity are $\pm45\%$ and 3.2$\times10^{-4}$~Jy~beam$^{-1}$, respectively. The inset at the bottom left corner shows the convolved beam of size $7\times7$~mas.}
\label{figcont1}
\end{figure}

In order to robustly examine the structure of the detected radio component {at 1.6~GHz}, we created images with different weighting schemes. Figure~\ref{figcont1} shows the 1.6~GHz contour image made with uniform weighting using a {\tt ROBUST} parameter of 0 in {\tt IMAGR}. {The {\it rms} noise in this image was $\sim7.0\times10^{-5}$\,Jy~beam$^{-1}$}. Figure~\ref{figcont2} shows a superimposition of the two radio images made with pure natural weighting (with {\tt ROBUST}=+5, in grey-scale), and pure uniform weighting (with {\tt ROBUST}=$-5$, in magenta-coloured contours). 
The {\it rms} noise in the latter image was $\sim1.3\times10^{-4}$\,Jy~beam$^{-1}$. All images have been convolved with a circular beam of size 7~mas, which was the best intermediate value.
The radio component is almost completely resolved out in the {purely uniformly-weighted} image. The brightest emission that remains at the source position is at the {$\sim3\sigma$} level. While its integrity needs to be examined with more sensitive VLBI observations in the future, we have shown this image to highlight its lack of centrally concentrated emission.

Assuming that three times the {\it rms} noise at 5~GHz {($=3.5\times10^{-5}$\,Jy~beam$^{-1}$)} was an upper limit to the (undetected) source flux density, {we derive the 1.6$-$5.0~GHz spectral index to be $\alpha=-1.5\pm0.5$, for a total source flux density of 650\,$\mu$Jy at 1.6~GHz. The {\it rms} noise at 1.6 and 5 GHz was used to estimate the spectral index error.} This spectral index value is an underestimate because the 1.6~GHz data probes shorter ($u, v$) spacings than the 5~GHz data and could pick up flux that is missed at 5~GHz. Steep spectrum radio cores are not uncommon in Seyfert galaxies \citep[e.g.,][]{Orienti10,Kharb10a,Bontempi12}. The steep radio spectrum rules out thermal or free-free emission as the radiation mechanism and instead supports a non-thermal origin.   

\begin{figure}[t]
\centerline{
\includegraphics[width=10cm,trim=1.8cm 0cm 0cm 0cm]{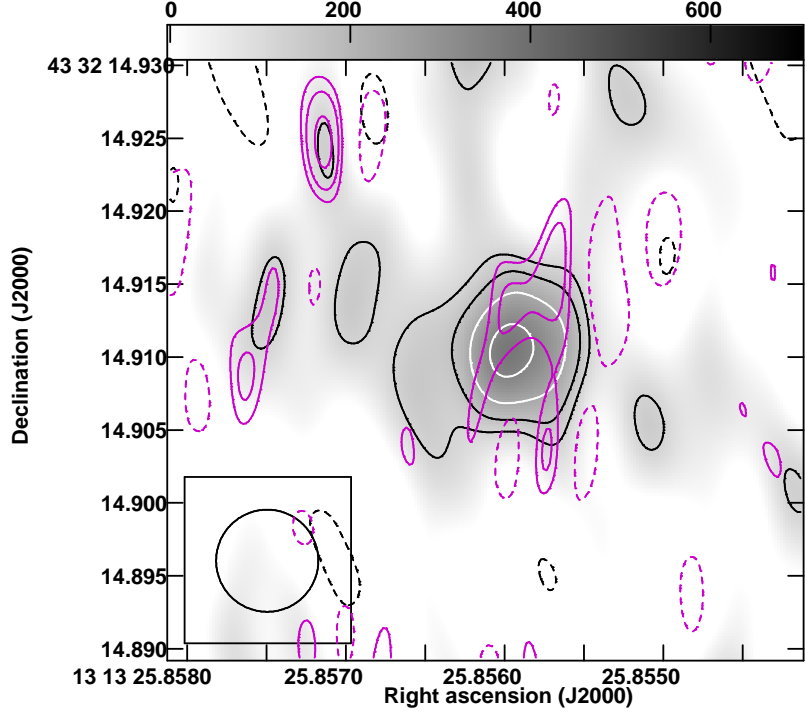}}
\caption{A superposition of the 1.6~GHz contour images made using pure natural weighting (in grey-scale) and pure uniform weighting (in magenta). The contours are in percentage of the peak intensity and increase in steps of $\sqrt{2}$: the lowest contour levels are $\pm32\%$ for both the images, and the peak intensities are 3.7$\times10^{-4}$~Jy~beam$^{-1}$ for the grey-scale and 6.7$\times10^{-4}$~Jy~beam$^{-1}$  for the magenta image. The convolved beam is of size $7\times7$ mas.}
\label{figcont2}
\end{figure}

The brightness temperature (in Kelvin) of the radio component at 1.6~GHz was estimated using the relation,

$\mathrm{T_{B}=1.8\times10^{9}~(1+z)~(\frac{S_{\nu}}{1\,mJy})\,(\frac{\nu}{1\,GHz})^{-2}~(\frac{\theta_{1}\theta_{2}}{mas^{2}})^{-1} }$,

\noindent
where $\theta_{1}, \theta_{2}$ are the major, minor axes of the best-fitting elliptical Gaussian for a resolved radio component \citep{Ulvestad05}. $T_B$ of the radio component is $\sim1.4\times10^7$~K. This suggests that the radio emission is {related to an AGN rather than a nuclear starburst}. Starbursts typically exhibit brightness temperatures of $<10^{5}$~K \citep{Condon91}. 

Assuming equipartition of energy between relativistic particles and the magnetic field \citep{Burbidge59}, we obtained the minimum {pressure, and the} particle energy (electrons and protons) at minimum pressure using the relations in \citet{OdeaOwen87},

$L_{rad}=1.2\times10^{27}~D_L^2~S~\nu^{-\alpha} (1+z)^{-(1+\alpha)} (\nu_u^{1+\alpha}-\nu_l^{1+\alpha}) (1+\alpha)^{-1}$,

$P_{min}=(2\pi)^{-3/7} (7/12) [c_{12} L_{rad} (1+k) (\phi V)^{-1}]^{4/7}$,

$B_{min}=[2 \pi (1+k) c_{12} L_{rad} (\phi V)^{-1}]^{2/7}$,

$E_{min}=[\phi V (2\pi)^{-1}]^{3/7} [L_{rad} (1+k) c_{12}]^{4/7}$,

\noindent
where $L_{rad}$ is the radio luminosity in erg~s$^{-1}$, $D_L$ is the luminosity distance in Mpc, $z$ is the source redshift, $S$ is the total flux density in Jy, $\nu$ is frequency in Hz, $\nu_u$ and $\nu_l$ are the upper and lower frequency cutoffs in Hz, respectively, $P_{min}$ is the minimum pressure in dynes~cm$^{-2}$, $k$ is the ratio of the relativistic proton to relativistic electron energy, $V$ is the source volume, $c_{12}$ is a constant depending on the spectral index and frequency cutoffs, $\phi$ is the volume filling factor, $B_{min}$ is the magnetic field at minimum pressure in G, and $E_{min}$ is the particle energy (electrons and protons) at minimum pressure in ergs. The total radio luminosity was estimated assuming that the radio spectrum extends from ($\nu_l$) 10~MHz to ($\nu_u$) 100~GHz with a spectral index of $\alpha=-1$. Furthermore, it was assumed that $k=1$. 
Table~\ref{tabequip} lists equipartition estimates for plasma filling factors of {unity and 0.5 \citep[e.g.,][]{Blustin09}.} 
The total energy in particles and fields, $E_{tot}$, is estimated as $E_{tot}=1.25\times E_{min}$, while the total energy density, $U_{tot}$, in erg~cm$^{-3}$ is $=E_{tot}(\phi V)^{-1}$.

The lifetime of electrons in the radio component undergoing both synchrotron radiative and inverse Compton losses on CMB photons was estimated using the relation \citep{vanderlaan69},

$t_{e}\simeq2.6\times10^{4} \frac{B^{1/2}}{B^{2}+B_{r}^{2}} \frac{1}{[(1+z)\nu_{b}]^{1/2}}$~yr,

\noindent
where $B$ was assumed to be the equipartition magnetic field $B_{min}$, and the break frequency $\nu_{b}$ was assumed to be 1.6~GHz. The magnetic field equivalent to the radiation, which was assumed to be predominantly CMB photons, $B_{r}$, was estimated using the relation, $B_{r}\simeq4\times10^{-6} (1+z)^{2}$~G. 
{The $B_{min}$ values of a few mG, log\,$U_{tot}$ of $\sim-5.0$ and electron lifetimes of $\sim$1000 yr are similar to estimates derived for the parsec-scale radio jets in Seyfert galaxies \citep[e.g.,][]{Kharb10a,Orienti10,Bontempi12,Kharb14a}. The significance of this finding will be discussed in Section~4.4.}

\subsection{Star formation rate}
While the high brightness temperature supports an AGN origin for the radio emission in KISSR\,1494, here we test the idea that the radio emission is star formation related. If we assume that all of the radio luminosity is attributable to star formation, a star formation rate (SFR) can be derived using the relation \citep{Condon92},

$\mathrm{SFR\,(M_\sun~yr^{-1})=L(W~Hz^{-1})/5.3\times10^{21}~\nu(GHz)^\alpha}$.

\noindent
We find that the upper limit to the star formation rate \citep[for stellar masses $\ge$~5~M$_\sun$; see][]{Condon92} from the {1.4~GHz luminosity ($=1.72\times10^{23}$~W~Hz$^{-1}$)} obtained in the kiloparsec-scale FIRST image is $\approx$43~M$_\sun$~yr$^{-1}$ for a spectral index of $\alpha=-0.8$. 

{The star formation rate can also be estimated from the narrow H$\alpha$ line. Adding the flux of the two narrow components of H$\alpha$ ($=2.9\times10^{-14}$~erg~cm$^{-2}$~s$^{-1}$) and following the relation in \citet{Kennicutt98},

$\mathrm{SFR\,(M_\sun~yr^{-1})=7.9\times10^{-42}~L_{[H\alpha]}(erg~s^{-1})}$,

\noindent
we derive a star formation rate of $\sim1.7$~M$_\sun$~yr$^{-1}$ for $\mathrm{L_{H\alpha}}=2.16\times10^{41}$~erg~s$^{-1}$.} 
The {SFR} in KISSR\,1494 derived from its optical spectrum therefore appears to be fairly low, and cannot account for the observed radio emission on kiloparsec-scales, which require the {SFR} to be $>40$~M$_\sun$~yr$^{-1}$. This finding supports the idea that the radio emission on kiloparsec-scales in KISSR\,1494 is AGN-outflow related.

\subsection{SMBH mass \& Eddington rate} 
Since equal peaked emission lines could arise in a rotating disk around the SMBH \citep{Smith12}, we use the suggestion of a disk to derive the dynamical mass of the SMBH in KISSR\,1494. Using the disk radius $r$ and the mean rotation speed $\langle{v}\rangle$, the mass of the SMBH can be derived as $M_{BH}~=~\frac{v^{2}~r}{G}$. We use the difference between the double peaks in the H$\alpha$ and H$\beta$ lines ($\Delta\lambda_{1}$ and $\Delta\lambda_{2}$) to determine $v$, where $2v~=~\frac{1}{2}[\frac{c\Delta\lambda_{1}}{6562.8} + \frac{c\Delta\lambda_{2}}{4861.36}]$, {and the denominators are the respective rest wavelengths in air.} 
{For $\Delta\lambda_{1}=6.15$ \AA\, and $\Delta\lambda_{2}=4.74$ \AA, we obtain $v=142.93$~km~s$^{-1}$. As the compact radio source has a dimension of $7.5\times5$~mas, we adopt an {\it arbitrary} disk diameter of 7.5~mas ($\sim8$ parsec). This yields a black hole mass of $M_{BH}=3.8\times10^{7}~M_{\sun}$ for KISSR\,1494.

The SMBH mass can also be estimated from the width and the luminosity of the broad component of the H$\alpha$ line (Figure~\ref{figspectra1}, Table~\ref{tabprop}). This method assumes that the broad component originates in the BLR and is not due to outflows \citep{Reines13}, and that the BLR clouds are in virial equilibrium with the SMBH \citep{Kaspi05}. Using the relation for black hole mass from \citet{Reines13} for a scale factor {($\epsilon$)} of unity:

$\mathrm{log\,(\frac{M_{BH}}{M_\sun})=6.57+0.47~log\,(L_{H\alpha})+2.06~log\,(FWHM_{H\alpha})}$,

\noindent
where H$\alpha$ line luminosity is in units of $10^{42}$~erg~s$^{-1}$ and FWHM in units of $10^3$~km~s$^{-1}$, we obtain $M_{BH}=3.7\times10^6~M_{\sun}$, for $\mathrm{L_{H\alpha}}=9.26\times10^{40}$~erg~s$^{-1}$. However, a broad component is also observed in both the [O {\sc iii}] $\lambda\lambda$4959, 5007 lines, which could be attributable to outflows related either to the AGN or nuclear star-formation \citep[e.g.,][]{Esquej14}. {If the broad H$\alpha$ line likewise had a significant outflow component, this black hole mass estimate would be unreliable.}

The bulge stellar velocity dispersion as derived by our line-fitting is $\sigma_\star=206.6\pm8.2$~km~s$^{-1}$. {Following the $M_{BH}-\sigma_\star$ relation for late-type galaxies \citep{McConnell13}:

$\mathrm{log(\frac{M_{BH}}{M_\sun})=8.07+5.06~log\,(\frac{\sigma_\star}{200~km~s^{-1}})}$,

\noindent
we derive a black hole mass, $M_{BH}=1.4\pm1.0\times10^8~M_{\sun}$, which is $\approx4$ times larger than the dynamical mass limit derived above. It is evident that the $M_{BH}-\sigma_\star$ relation has a lot of scatter when it comes to late-type galaxies.}

{All three estimates of black hole mass are very crude. For the lack of a more reliable measurement, we have adopted the BH mass derived from the $M_{BH}-\sigma_\star$ relation in the rest of the paper.} In Table~\ref{bhmass} we list all the BH mass estimates along with their caveats.

We use the combined flux of the narrow components of the [O {\sc iii}] $\lambda5008$ line ($=1.21\times10^{41}$~erg~s$^{-1}$) to estimate the bolometric luminosity ($L_{bol}$) using the relation from \citet{Heckman04}: $L_{bol}/L_{5000}\approx3500$, where $L_{5000}$ is the monochromatic continuum luminosity at a 5000 \AA\, rest frame. We derive $L_{bol}=4.2\times10^{44}$~erg~s$^{-1}$. 
{For a black hole mass of $1.4\times10^8~M_{\sun}$, the Eddington luminosity ($\mathrm{\equiv1.25\times10^{38}~M_{BH}/M_\sun}$) is $\approx1.7\times10^{46}$~erg~s$^{-1}$ 
and the Eddington rate ($\mathrm{\equiv L_{bol}/L_{Edd})~is \sim0.02}$.}
The accretion rate in KISSR\,1494 seems to be typical of Seyfert galaxies in the literature \citep[e.g.,][]{Ho08}. In short, the double peaked emission line {Seyfert} KISSR\,1494, does not distinguish itself from other Seyfert galaxies in terms of its  black hole mass and accretion properties.

\section{Discussion}
\label{secdisc}
A single steep spectrum radio component is detected in the {double peaked emission line Seyfert galaxy}, KISSR\,1494. The spectrum does not favour thermal free-free emission, but rather optically thin synchrotron emission. While steep spectrum radio cores are fairly prevalent in Seyfert galaxies, they may represent a different phenomena compared to the typical flat/inverted spectra radio cores observed in radio-loud sources and even many Seyfert galaxies, that are traditionally regarded as the synchrotron-self-absorbed unresolved bases of radio jets. The high brightness temperature of $T_B\sim1.4\times10^7$~K also favours a non-thermal AGN-related origin rather than a nuclear starburst which is expected to have $T_B<10^5$~K. Additionally, the low star-formation rate derived from the optical spectrum {(SFR~$\sim1.7$~M$_\sun$~yr$^{-1}$)}, does not support the idea of the kiloparsec-scale radio emission in KISSR\,1494 being starburst related, which requires an SFR $>40$~M$_\sun$~yr$^{-1}$. {We discuss below the favoured mechanisms for explaining double peaked emission lines in AGN in the context of high resolution radio observations of KISSR\,1494}.

\subsection{The Binary SMBH Scenario}
{The presence of a single radio component {\it per se} does not rule out the binary black hole scenario. One possibility could be that} the two black holes have separations less than the size of the detected radio component, i.e., if they are separated by $<8$~parsec. However in this case, the NLR clouds giving rise to the double peaks will have to be restricted to those spatial scales as well, making this an unattractive proposition. 
{The NLR clouds typically exist on spatial scales of tens to hundreds of parsecs from the central engine  \citep[e.g.,][]{Crenshaw00}. In addition, it is unclear what media (equivalent to dusty tori around BLR clouds) could shield the  spatially extended NLR clouds of the two black holes from mixing and losing their individual identities.}

An alternate scenario could be that a second black hole is indeed present, but has associated radio emission that is below the {\it rms} noise in the final image ($\sim6.5\times10^{-5}$~Jy~beam$^{-1}$ at 1.6~GHz). There is a well established correlation between the core radio luminosity at 5~GHz ($L_5$, W~Hz$^{-1}$) and the BH mass for nearby early-type galaxies \citep{Franceschini98}: $L_5\propto M_{BH}^{2.73}$. What is more relevant to KISSR\,1494 is that for a handful of late-type galaxies considered in this study, the BH mass exponent is still close to $2.7$. Assuming the 1.6$-$5.0~GHz core spectral index to be $-1$, we estimate that the upper limit to the radio luminosity at 5~GHz from the second BH would be $\sim1.5\times10^{20}$~W~Hz$^{-1}$. From the $L_5 - M_{BH}$ relation, this implies that the mass of the second BH would be $\lesssim2.5\times10^7~M_\sun$. 
{At the upper limit, this black hole mass lies close to the $M_{BH}-\sigma_\star$ mass of the primary black hole (that emits at radio frequencies) within error limits. It would be surprising though if a major merger took place in KISSR\,1494 in its past and left no tell-tale signatures of a disturbance, like tidal tails, shells or loops.} In addition, this merger apparently left the spiral features of the host galaxy intact. Major mergers of spiral galaxies are expected to lead to the formation of elliptical galaxies \citep[e.g.,][]{Schweizer82}. This disfavours the suggestion of a second black hole with a mass nearly equal to the primary black hole (with the associated radio emission), but having only faint (undetectable) radio emission. It is also instructive to remember that \citet{BurkeSpolaor11} found only two binary AGN in a sample of 3114 radio-bright AGN observed with VLBI, attesting to the rarity of sources {with binary back holes}.

An important caveat to the discussion above must be mentioned. {It is possible that the detected radio component at 1.6~GHz is not coincident with the position of the central supermassive black hole, but is instead a bright portion of a radio outflow, like a compact jet knot or a ``hot spot".} However, it is not possible to verify this with the present day optical telescopes. Only a more sensitive VLBI study at multiple wavelengths may detect either more radio components from the radio outflow or reveal the spectral index information with greater confidence, and thereby elucidate the true nature of the detected radio feature. For the purposes of this paper, we have assumed that the detected radio component is indeed coincident with the supermassive black hole in KISSR\,1494.

\subsection{The Rotating Disk Scenario}
{\citet{Smith12} have proposed that equal-peaked AGN, of which KISSR\,1494 is an example, could have their line emission originating in a rotating ring with relatively little gas at small radii. The fact that the radio component detected at 1.6~GHz in KISSR\,1494} is not compact and gets completely resolved out in some image weighting schemes {(see Figure~\ref{figcont2})} can be reconciled with the fact that the radio emission is from a disk or a ring/torus. Could the same rotating disk or ring/torus give rise to the double peaked emission lines~? This is unlikely because of the small size of the radio feature ($\sim8\times6$~parsec) $-$ the NLR clouds are expected to be distributed on much larger $10-100$ parsec-scales, extending beyond the parsec-scale torus. However, it is possible that the disk is much more extended than the region probed by {VLBI} and the NLR clouds form part of the outer regions of this disk. {The missing diffuse radio emission going from kiloparsec to parsec-scales could also arise in an extended disk emitting at radio frequencies.} \citet{Mulchaey96} have demonstrated that if the NLR is in a disk then the V-shaped ionisation cones will always be observed in Seyfert galaxies, irrespective of our line of sight being inside or outside of the cones. The prevalence of ionisation cones therefore favours NLR disks. 

\citet{Gallimore04} have suggested that the flat/inverted spectrum radio core in the Seyfert galaxy NGC\,1068 is thermal bremsstrahlung (free-free) emission from the inner hot ionised edge of the torus. \citet{Roy98} have discussed the possibility that the radio core in NGC\,1068 is direct synchrotron emission from the torus instead. 
{The steep spectrum along with the moderate brightness temperature $T_B\sim10^7$~K in the radio core of KISSR\,1494} does not favour synchrotron self-absorption from the torus as an explanation, but rather optically thin synchrotron emission. 
For synchrotron self-absorption to occur, $T_B>10^9$~K is typically required. Steep-spectrum parsec-scale radio emission from another Seyfert galaxy, NGC\,4388, has been suggested to originate in a disk \citep{Giroletti09}. The size of the radio emitting disk in NGC\,1068 is $\sim$0.4 parsec with a $T_B$ = $2.6\times10^6$~K, while the size of the radio disk in NGC\,4388 is $\sim$0.5 parsec with a $T_B$ = $1.3\times10^6$~K. 
The NLR in NGC\,1068 and NGC\,4388 are probably not disky because their SDSS spectra do not exhibit double-peaked emission lines, unless their NLR disks are nearly face-on. 
If the radio component in KISSR\,1494 is indeed emission from a disk, its size is at least ten times larger than disks in NGC\,1068 and NGC\,4388. Its brightness temperature is five to ten times higher as well. The significance of this difference is unclear at present. 

\subsection{The Radio Outflow Scenario}
If a radio outflow impacts the gas clouds in the NLR, then the approaching and receding emission-line gas clouds could give rise to the double peaked emission line spectra. The lack of a clear {parsec-scale} core-jet structure is not consistent with a powerful radio outflow in KISSR\,1494. However, a broad wind-like {parsec-scale} radio outflow cannot be ruled out. 

As noted in Section~\ref{secintro}, the total radio flux density of KISSR\,1494 with the VLA FIRST survey at 1.4~GHz is $\sim$23~mJy. This flux density drops to 0.65~mJy at 1.6~GHz with VLBI (present study). Hence most of the radio flux density comes from a diffuse radio component on scales between a few arcseconds and 10 {mas}. AGN-driven outflows have indeed been observed on scales of a few kiloparsec in Seyfert galaxies, especially {with sensitive radio observations} \citep[e.g.,][]{Kharb06,Gallimore06,Kharb14b}. If we assume that {the kiloparsec-scale emission is lobe emission from a} broad AGN outflow in KISSR\,1494, then we can estimate the time-averaged kinetic power ($Q$) of this outflow following the relations in \citet{Punsly11} for radio-powerful sources \citep[see also][]{Willott99}:

$\mathrm{Q\approx1.1\times10^{45}~[(1+z)^{1+\alpha}~Z^2~F_{151}]^{6/7}~erg~s^{-1}}$,

$\mathrm{Z\equiv 3.31 - 3.65 [(1+z)^4 - 0.203 (1+z)^3 }$ \\ 
$\mathrm{+ 0.749 (1+z)^2 + 0.444 (1+z) + 0.205]^{-0.125}}$,
  
\noindent
where $F_{151}$ is the 151~MHz flux density {and $z$ is the source redshift. We derive $F_{151}$ using the 1.4~GHz flux density from FIRST and a lobe spectral index of $\alpha=-0.8$, and obtain a kinetic power of $Q\approx1.5\times10^{42}$~erg~s$^{-1}$. This $Q$ value is typical} of outflows in low-luminosity AGN \citep[e.g.,][]{Mezcua14}. 

\citet{Blundell07} have suggested that the radio core emission in radio-quiet quasars is thermal free-free emission from a slow dense disk wind. The radio data of KISSR\,1494 are consistent with a disk wind that emits synchrotron emission instead. It could indeed be wind emission from the tenuous corona region above the accretion disk, which has a sufficient supply of relativistic electrons and strong magnetic fields \citep[e.g.,][]{Merloni02,Laor08}. In this scenario, the radio emission does come from a less collimated outflow compared to a relativistic jet. 
{As mentioned by \citet{Marscher06}, the corona itself could be the low altitude portion of a wind flowing from the accretion disk.}
The double-peaked emission lines in the source could also be explained if the NLR clouds are a part of this broader outflow. Even the equal peaks of the H$\alpha$ and H$\beta$ emission lines can be explained if they arise in a highly symmetric outflow close to the central engine, whose intensities are not significantly affected by dust obscuration. 

\subsection{Disk versus Outflow}
We conclude that the radio continuum emission and the double-peaked emission lines in the spectrum of KISSR\,1494 are consistent {primarily} with two explanations: [A] the mas-scale radio emission is the parsec-scale base of a synchrotron wind from the magnetised corona above the accretion disk. The NLR clouds are also a part of this outflow and extend to larger scales, probably of the order of tens to hundreds of parsecs \citep[e.g.,][]{Crenshaw00,Ruiz05}, and [B] the mas-scale radio emission comes from the (static, as in not outflowing) inner ionised edge of the accretion disk or torus; this disk extends to tens or hundreds of parsecs and also incorporates the rotating NLR clouds. In both these cases
{the radio emission is from the synchrotron mechanism.} While the extensive literature on the presence of parsec-scale radio outflows in Seyferts \citep[e.g.,][]{Giroletti09} supports explanation [A], the fact that KISSR\,1494 is an ``equal-peaked" AGN supports the disk explanation [B]. {However, as mentioned in Section 4.3,} a highly symmetric {AGN-driven NLR} outflow where the line intensities from the receding flow are not significantly obscured by dust, could also support explanation [A] {in ``equal-peaked'' AGN}. Therefore, although our present data cannot clearly distinguish between explanations [A] and [B], there appears to be more supporting evidence for explanation [A], i.e., a non-thermal coronal disk wind. 
{Finally, the radio and NLR emission may be decoupled. The radio emission may come from an outflow while the NLR may be in a disk, and {\it vice versa}.}

If the radio core in KISSR\,1494 is indeed an outflowing coronal wind, it is interesting to note that its properties (magnetic field strength, total energy density, electron lifetime, kinetic power) are no different from the radio ``jet'' properties derived in other Seyfert galaxies. This supports the idea that radio ``jets'' and outflowing coronal winds are indistinguishable in Seyferts. The implication of this finding needs to be explored further.

\section{Summary and Conclusions}
We have carried out phase-referenced VLBI observations at 1.6 and 5~GHz of the double emission line peaked Seyfert galaxy, KISSR\,1494. The primary findings are:

\begin{enumerate}
\item
A single radio component of size $7.5\times5.1$~mas ($\sim8\times6$ parsec) is observed at 1.6~GHz. There is no detection at 5~GHz. This implies a radio spectral index steeper than $-1.5\pm0.5$. The high brightness temperature of the radio component ($T_B\sim1.4\times10^7$~K) and the steep radio spectrum suggest that it is non-thermal synchrotron emission, rather than thermal free-free emission. 
\item
The star formation rate derived from the {H$\alpha$ line in the optical spectrum ($\sim1.7$~M$_\sun$~yr$^{-1}$)} is too low to account for the radio luminosity on kiloparsec-scales which requires a star formation rate greater than $\sim40$~M$_\sun$~yr$^{-1}$. This supports the idea that the radio emission is AGN-outflow related.
\item
{The presence of a single radio core could {\it per se} be construed as being inconsistent with} the binary black hole picture for producing double peaked emission lines in this galaxy. {However, the  suggestions that binary black holes exist within the $\sim$8 parsec region of the detected radio component, or that one of the black holes is ``radio-quiet", could still hold. These suggestions are not without their own characteristic difficulties.}
\item The twin peaks observed in H$\alpha$, H$\beta$ and [S {\sc ii}] are identical, suggesting that the NLR clouds could be part of a rotating disk. That the radio component gets completely resolved out in certain image weighting schemes can be reconciled with the fact that the radio emission is from a disk or a torus. 
\item
The lack of a clear core-jet structure is inconsistent with a strong, collimated radio outflow being the primary mechanism for the production of the double emission line peaks. However, a broad outflow arising in an accretion disk corona is favoured by our data. 
\item
From the $M_{BH}-\sigma_{\star}$ relation {for late-type galaxies, a black hole mass of $M_{BH}=1.4\pm1.0\times10^8~M_{\sun}$ is derived, accreting at an Eddington rate of $\sim0.02$.} In terms of its black hole mass and accretion properties, as well as the magnetic field strength and ``minimum'' total energy (charged particles and magnetic fields) derived under the equipartition assumption, {the double peaked emission line Seyfert KISSR\,1494,} does not distinguish itself from other Seyfert galaxies. 
\item
{To summarise, the VLBI data are consistent with two explanations:} the radio emission is the parsec-scale base of a synchrotron wind from the magnetised corona above the accretion disk, or the radio emission comes from the inner ionised edge of the accretion disk or torus that is not outflowing. In the first case, the NLR clouds {could also be} a part of this broad outflow, while in the latter, the NLR clouds {could} form a part of an extended disk beyond the torus. {It is also possible that the radio emission comes from an outflow while the NLR clouds are in a disk, or {\it vice versa}.} While with the present data, it is not possible to clearly distinguish between these scenarios, there appears to be greater circumstantial evidence supporting the coronal wind picture in KISSR\,1494. From the kiloparsec-scale radio emission, the time-averaged kinetic power of this outflow is estimated to be $Q\approx1.5\times10^{42}$~erg~s$^{-1}$, a value typical of radio outflows in low-luminosity AGN. {This suggests that radio ``jets'' and outflowing coronal winds are indistinguishable in Seyferts.}
\end{enumerate}

\acknowledgments
We thank the anonymous referee for providing useful suggestions which have improved this manuscript.
MD and PK would like to thank S. Ramya for valuable advice regarding the spectral analysis. ZP acknowledges financial support from the International Space Science Institute (ISSI). The National Radio Astronomy Observatory is a facility of the National Science Foundation operated under cooperative agreement by Associated Universities, Inc. This research has made use of the NASA/IPAC Extragalactic Database (NED) which is operated by the Jet Propulsion Laboratory, California Institute of Technology, under contract with the National Aeronautics and Space Administration. Funding for the SDSS and SDSS-II has been provided by the Alfred P. Sloan Foundation, the Participating Institutions, the National Science Foundation, the U.S. Department of Energy, the National Aeronautics and Space Administration, the Japanese Monbukagakusho, the Max Planck Society, and the Higher Education Funding Council for England. 

{\it Facilities:} \facility{VLBA}, \facility{Sloan}

\bibliographystyle{apj}
\bibliography{ms}

\onecolumn
\begin{table}
\caption{Fitted Line Parameters}
\begin{center}
\begin{tabular}{lcccccc}
\hline\hline
{Line}&{$\lambda_{0}$}&{$\lambda_{c}\pm$error}&{$\Delta\lambda\pm$error}&{$f_{p}\pm$error}& {log $f$}&{log $L$}\\
{(1)} & {(2)} & {(3)} & {(4)} & {(5)} & {(6)} & {(7)}\\
\hline
$[\mathrm {SII}]$  & 6718.3   &6716.36$\pm$0.33 & 2.91$\pm$0.08  &51.85$\pm$3.46  & $-$14.42 &    40.45 \\  
                   &          &6722.16$\pm$0.25 & 2.26$\pm$0.18  &43.43$\pm$2.91  & $-$14.61 &    40.26 \\  
$[\mathrm {SII}]$  & 6732.7   &6730.36$\pm$0.33 & 2.92$\pm$0.08  &38.75$\pm$2.67  & $-$14.55 &    40.33 \\  
                   &          &6736.17$\pm$0.25 & 2.27$\pm$0.18  &35.36$\pm$2.32  & $-$14.70 &    40.18 \\  
$[\mathrm {NII}]$  & 6549.9   &6548.25$\pm$0.22 & 2.84$\pm$0.08  &75.26$\pm$3.96  & $-$14.27 &    40.60 \\  
                   &          &6553.84$\pm$0.22 & 2.21$\pm$0.18  &52.69$\pm$2.29  & $-$14.54 &    40.34 \\  
$[\mathrm {NII}]$  & 6585.3   &6583.26$\pm$0.22 & 2.86$\pm$0.08  &222.01$\pm$3.96 & $-$13.80 &    41.08 \\  
                   &          &6588.87$\pm$0.22 & 2.22$\pm$0.18  &155.44$\pm$2.29 & $-$14.06 &    40.81 \\  
H$\alpha$          & 6564.6   &6561.99$\pm$0.26 & 2.85$\pm$0.08  &232.45$\pm$13.17& $-$13.78 &    41.09 \\  
                   &          &6568.14$\pm$0.23 & 2.21$\pm$0.18  &222.08$\pm$5.81 & $-$13.91 &    40.96 \\  
                   &          &6570.23$\pm$0.65 & 16.19$\pm$0.58 &30.52$\pm$3.04  & $-$13.91 &    40.97 \\  
H$\beta$           & 4862.7   &4860.37$\pm$0.22 & 2.11$\pm$0.08  &52.59$\pm$3.00  & $-$14.55 &    40.32 \\  
                   &          &4865.11$\pm$0.17 & 1.64$\pm$0.18  &51.55$\pm$2.66  & $-$14.67 &    40.20 \\  
$[\mathrm {OIII}]$ & 4960.3   &4963.02$\pm$2.66 & 2.47$\pm$1.64  &38.20$\pm$58.23 & $-$14.63 &    40.25 \\  
                   &          &4959.20$\pm$4.72 & 2.92$\pm$0.58  &42.12$\pm$72.03 & $-$14.51 &    40.36 \\  
                   &          &4958.36$\pm$1.75 & 5.60$\pm$0.88  &10.15$\pm$7.79  & $-$14.85 &    40.03 \\  
$[\mathrm {OIII}]$ & 5008.2   &5011.00$\pm$2.66 & 2.49$\pm$1.64  &112.70$\pm$58.23& $-$14.15 &    40.72 \\  
                   &          &5007.13$\pm$4.72 & 2.95$\pm$0.58  &124.26$\pm$72.03& $-$14.04 &    40.84 \\  
                   &          &5006.04$\pm$1.76 & 5.65$\pm$0.88  &29.96$\pm$7.79  & $-$14.37 &    40.50 \\  
$[\mathrm {OI}]$   & 6302.0   &6300.45$\pm$2.81 & 2.41$\pm$0.40  &13.18$\pm$14.36 & $-$15.10 &    39.78 \\  
                   &          &6305.32$\pm$3.13 & 2.56$\pm$2.31  &11.44$\pm$4.44  & $-$15.13 &    39.74 \\  
$[\mathrm {OI}]$   & 6365.5   &6364.45$\pm$2.81 & 2.84$\pm$0.94  &4.39$\pm$14.36  & $-$15.51 &    39.37 \\  
                   &          &6369.37$\pm$3.13 & 2.25$\pm$2.74  &3.81$\pm$4.44   & $-$15.67 &    39.21 \\  
$[\mathrm {OII}]$  & 3727.1   &3729.02$\pm$1.01 & 1.62$\pm$0.08  &48.40$\pm$53.66 & $-$14.71 &    40.17 \\  
                   &          &3732.01$\pm$0.44 & 1.26$\pm$0.18  &44.76$\pm$23.89 & $-$14.85 &    40.02 \\  
                   &          &3727.81$\pm$1.67 & 2.67$\pm$0.32  &72.88$\pm$37.55 & $-$14.31 &    40.56 \\  
H$\gamma$          & 4341.7   &4338.60$\pm$0.38 & 1.88$\pm$0.08  &16.03$\pm$1.81  & $-$15.12 &    39.75 \\  
                   &          &4343.22$\pm$0.24 & 1.46$\pm$0.18  &19.10$\pm$2.12  & $-$15.16 &    39.72 \\  
\hline
\end{tabular}
\end{center}
{Column~1: Emission lines that were fitted with Gaussian components. {Column~2: Rest wavelength in vacuum in $\AA$.} Columns~3, 4: Central wavelength and line width in $\AA$ along with respective errors. Column~5: Peak line flux in units of $10^{-17}$~erg~cm$^{-2}$~s$^{-1}~\AA^{-1}$ with error. Column~6: Logarithm of the total line flux in erg~cm$^{-2}$~s$^{-1}$. Column~7: Logarithm of the line luminosity in erg~s$^{-1}$.}
\label{tabprop}
\end{table}

\begin{table}
\caption{Equipartition Estimates}
\begin{center}
\begin{tabular}{cccccccc}
\hline\hline
{$L_{rad}$}              &{$\phi$} &   {$P_{min}$}          & {$E_{min}$}            & {$B_{min}$}& {$E_{tot}$}               & {$U_{tot}$} & {$t_e$}\\
\hline
$1.4\times10^{41}$ & 0.5       & $5.2\times10^{-6}$ & $3.2\times10^{52}$ & 7.4              & $4.0\times10^{52}$ & $1.1\times10^{-5}$ & 980 \\
$1.4\times10^{41}$ & 1.0       & $3.5\times10^{-6}$ & $4.3\times10^{52}$ & 6.1              & $5.4\times10^{52}$ & $7.4\times10^{-6}$ & 1320  \\
\hline
\end{tabular}
\end{center}
{Column~1: Total radio luminosity in erg~s$^{-1}$. Column~2: Plasma filling factor. Column~3: Minimum pressure in dynes~cm$^{-2}$.
Column~4: Minimum energy in ergs. Column~5: Minimum B-field in mG. Column~6: Total energy in particles and fields, $E_{tot}$ ($=1.25\times E_{min}$) in ergs. Column~7: Total energy density, $U_{tot}=E_{tot}(\phi V)^{-1}$ in erg~cm$^{-3}$. {Column~8: Electron lifetimes in yr. See Section~3.1 for details.}}
\label{tabequip}
\end{table}

\begin{table}
\caption{Black Hole Mass Estimates}
\begin{center}
\begin{tabular}{lcl}
\hline\hline
{Method}&{$M_{BH}$}&{Caveat}\\
\hline
Dynamical                     & $3.8\times10^7M_\sun$ & Arbitrary disk-size from radio data\\
$M_{BH}-\sigma_\star$& $1.4\times10^8M_\sun$ & Large scatter in relation for late-type galaxies\\
Broad H$\alpha$          & $3.7\times10^6M_\sun$ & Broad component could signify outflow\\
\hline
\end{tabular}
\end{center}
{}
\label{bhmass}
\end{table}

\end{document}